# OpenCCM : une infrastructure à composants pour le déploiement d'applications à base de composants CORBA

Frédéric Briclet, Christophe Contreras et Philippe Merle

*Projet Jacquard INRIA Futurs*
*Laboratoire d'Informatique Fondamentale de Lille (LIFL)*
*UMR CNRS 8022 / Université des Sciences et Technologies de Lille (USTL)*
*59655 Villeneuve d'Ascq Cedex, France*
*Frederic.Briclet@lifl.fr, Christophe.Contreras@lifl.fr, Philippe.Merle@inria.fr*

RÉSUMÉ. *Le déploiement de composants logiciels pour la construction d'applications réparties consiste à coordonner un ensemble de tâches élémentaires comme le téléchargement des binaires sur les sites d'exécution, leur chargement en mémoire, la création d'instances de composants, l'interconnexion de leurs ports, la configuration de leurs propriétés métiers et techniques. Automatiser le processus de déploiement nécessite alors la présence d'une infrastructure logicielle elle-même répartie sur les différents sites d'exécution.*

*Cet article présente les caractéristiques d'une telle infrastructure pour le déploiement d'applications à base de composants CORBA. Cette dernière a été conçue et réalisée dans le cadre de notre plate-forme OpenCCM, une implantation libre du modèle de composants CORBA. La principale caractéristique de cette infrastructure est qu'elle est elle-même conçue sous la forme d'assemblages de composants CORBA. Ceci autorise son assemblage dynamique lors de son déploiement sur les sites d'exécution.*

ABSTRACT. *Deployment of software components for building distributed applications consists of the coordination of a set of basic tasks like uploading component binaries to the execution sites, loading them in memory, instantiating components, interconnecting their ports, setting their business and technical attributes. The automation of the deployment process then requires the presence of a software infrastructure distributed itself on the different execution sites.*

*This paper presents the characteristics of such an infrastructure for the deployment of CORBA component-based applications. This latter is designed and implemented in the context of our OpenCCM platform, an open source implementation of the CORBA Component Model. The main characteristic lays on the fact that this infrastructure is itself designed as a set of CORBA component assemblies. This allows its dynamic assembly during its deployment over the execution sites.*

MOTS-CLÉS : *déploiement, composant, intergiciel, CORBA.*

KEY WORDS: *deployment, component, middleware, CORBA.*

**1. Introduction**

Depuis quelques années, la notion de déploiement logiciel évolue rapidement, en particulier avec l'interconnexion croissante des applications sur les réseaux de communication. Autrefois cantonné à de simples activités de copie et d'installation poste à poste, le déploiement devient désormais une préoccupation majeure dans le cadre des applications construites avec des technologies à composants [SZY 02] telles que les Enterprise Java Beans (EJB) de SUN Microsystems, le modèle de composants CORBA (CCM) de l'Object Management Group (OMG), le canevas .NET de Microsoft ou le modèle Fractal [BRU 04] du consortium ObjectWeb.

Le déploiement de composants logiciels pour la construction d'applications réparties nécessite de coordonner un ensemble de tâches élémentaires comme le téléchargement des binaires de composants sur leur site d'exécution, le chargement de ces binaires dans la mémoire des serveurs d'applications, l'instanciation des binaires pour créer des instances de composants, la configuration des attributs métiers des instances de composants, l'établissement des liaisons entre les interfaces requises et offertes des différents composants, puis le démarrage effectif des instances de composants. Automatiser la coordination du processus de déploiement requiert la présence d'une infrastructure logicielle elle-même répartie sur les différents sites d'exécution. Cette dernière doit offrir les primitives élémentaires de contrôle du téléchargement des binaires de composants, d'instanciation et de configuration des instances de composant ainsi que les procédures de coordination de l'exécution distribuée de ces primitives.

Cet article présente les caractéristiques d'une telle infrastructure pour le déploiement d'applications à base de composants CORBA. Cette dernière, nommée *Distributed Computing Infrastructure* (DCI), a été conçue et réalisée dans le cadre de notre plate-forme OpenCCM [MAR 01], une implantation libre du modèle de composants CORBA [OPE 02]. La principale caractéristique de l'infrastructure DCI est qu'elle est elle-même conçue sous la forme d'assemblages de composants CORBA. Ces différents composants sont déployés et assemblés dynamiquement afin de constituer l'infrastructure DCI répartie sur les sites d'exécution. Une approche similaire de conception est aussi décrite dans les travaux de Ayed et al. [AYE 04].

La suite de cet article est organisée comme suit. La section 2 présente les caractéristiques essentielles du déploiement définies dans la spécification du modèle de composants CORBA. La section 3 décrit l'architecture et chacun des composants de l'infrastructure de déploiement de la plate-forme OpenCCM. Enfin, la section 4 conclut et dégage de nombreuses perspectives de travail.

**2. Déploiement de composants CORBA**

La troisième version de la spécification CORBA introduit un modèle de composants logiciels [OMG 02] enrichissant l'architecture orientée objet de CORBA [WAN 01]. Elle spécifie en particulier les interfaces de programmation de l'infrastructure de déploiement et le scénario type du processus de déploiement d'applications construites à base de composants CORBA.

## 2.1. *Les interfaces de déploiement du modèle de composants CORBA*

Le modèle de déploiement CORBA définit six interfaces modélisant une infrastructure de déploiement CCM (voir Figure 1) :

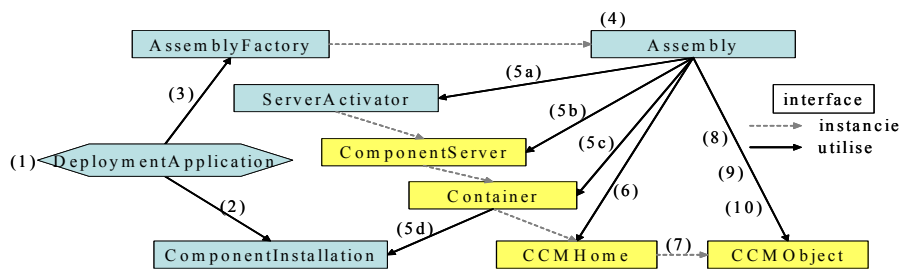

**Figure 1.** *L'architecture de déploiement du modèle de composants CORBA*

– L'interface *ComponentInstallation* modélise un gestionnaire d'archives de composants et fournit les opérations pour installer, retrouver et désinstaller les archives et implantations de composants sur les nœuds de l'infrastructure. Il ne doit y avoir qu'une instance de *ComponentInstallation* par nœud.

– L'interface *AssemblyFactory* modélise le gestionnaire d'instances d'assemblages et fournit les opérations pour créer des assemblages à partir de leurs archives, retrouver et détruire les instances d'assemblage.

– L'interface *Assembly* représente une instance d'assemblage déployée sur l'infrastructure. Cette interface permet de contrôler l'instanciation (ou déploiement) de l'assemblage ainsi que sa destruction (ou repliement).

– L'interface *ServerActivator* représente le gestionnaire de serveurs d'applications et fournit les opérations pour créer, retrouver et détruire des serveurs. Il ne doit y avoir qu'une instance de *ServerActivator* par nœud.

– L'interface *ComponentServer* représente un serveur d'applications et fournit les opérations pour créer, retrouver et détruire des conteneurs.

– L'interface *Container* représente un conteneur pouvant accueillir des maisons de composants et fournit les opérations pour installer, retrouver et détruire ces maisons.

En complément à ces six interfaces, l'infrastructure de déploiement utilise deux interfaces héritées respectivement par toutes les instances de maisons et de composants. L'interface *CCMHome* fournit des opérations de création, de recherche et de destruction d'instances de composant, et l'interface *CCMObject* fournit des opérations pour la configuration des instances, en particulier l'introspection, la connexion et la déconnexion des ports.

### 2.2. Le scénario type du processus de déploiement

Le scénario type d'un déploiement d'assemblages de composants CORBA est le suivant (se reporter à la figure 1 pour les numéros). L'utilisateur fournit une archive d'assemblage à déployer sur l'infrastructure via l'application de déploiement (1). L'application de déploiement contacte les interfaces *ComponentInstallation* afin de télécharger sur chaque site les archives de composants nécessaires (2). Puis elle contacte l'interface *AssemblyFactory* pour créer une instance d'assemblage (3). Le lien URL passé en paramètre pointe sur l'archive que doit télécharger l'instance d'assemblage. L'application de déploiement contacte l'interface *Assembly* pour démarrer le processus de déploiement de l'assemblage (4). L'instance d'assemblage contacte ensuite successivement les interfaces *ServerActivator*, *ComponentServer* et *Container* sur chaque site afin de créer respectivement les différents serveurs d'applications (5a) et conteneurs (5b) puis installer les maisons de composants (5c) nécessaires à l'assemblage. Les conteneurs obtiennent les archives des composants via l'interface du *ComponentInstallation* local (5d). L'instance d'assemblage configure les attributs des instances de maisons précédemment installées (6). Puis elle ordonne aux maisons la création des instances de composants conformément à l'information du descripteur d'assemblage (7) et configure les attributs de ces instances (8). L'instance d'assemblage interconnecte les ports des instances de composants via les opérations de connexion fournies par l'interface *CCMObject* (9). L'instance d'assemblage indique alors à chaque instance de composant la fin de sa configuration impliquant son démarrage effectif (10). Enfin, l'instance d'assemblage enregistre, si nécessaire, les références des maisons et des composants dans des services d'annuaires.

### 2.3. Critiques sur l'architecture de déploiement CCM

Le modèle de composants CORBA définit précisément les différentes étapes du processus de déploiement d'applications à base de composants CORBA. On peut aussi constater que l'instance d'assemblage est au cœur du processus de déploiement et agit comme le coordinateur des différentes tâches élémentaires du déploiement. Cependant, l'architecture de déploiement CCM ne spécifie pas explicitement les dépendances entre les différentes entités. Le standard ne précise pas les moyens mis en œuvre pour obtenir les références des *ComponentInstallation* et de l'*AssemblyFactory* par l'application de déploiement, celles des *ServerActivator* et des services d'annuaires par l'assemblage et celle du *ComponentInstallation* local par le conteneur. Les moyens mis en oeuvre sont inhérents aux choix d'implantation d'une infrastructure de déploiement CCM. Dans notre infrastructure OpenCCM DCI, toutes les entités participant au déploiement sont implantées sous la forme de composants CORBA et leurs dépendances sont explicitement représentées par des réceptacles, c'est-à-dire les interfaces requises par les composants.

## 3. L'infrastructure de déploiement OpenCCM DCI

Cette section présente l'infrastructure de déploiement OpenCCM DCI. Celle-ci a été conçue, spécifiée et réalisée dans le cadre du projet européen IST COACH [HOF 03a, HOF 03b]. L'architecture, les principes de conception et les sept composants de l'infrastructure DCI sont décrits dans les sections suivantes.

### 3.1. L'architecture de l'infrastructure DCI

L'infrastructure OpenCCM DCI est une application répartie sur différents sites du réseau. Son architecture logicielle est décomposée en trois couches principales : le domaine de déploiement, la machine de déploiement répartie et les noeuds d'accueil (voir Figure 2).

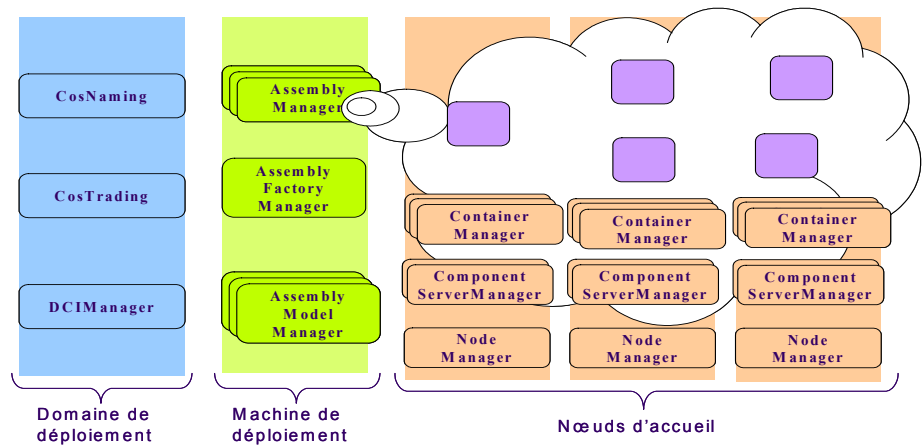

**Figure 2.** *L'architecture de l'infrastructure de déploiement OpenCCM DCI*

Le ***domaine de déploiement*** regroupe l'ensemble des services communs à savoir les services d'annuaires *CosNaming* et *CosTrading* et le gestionnaire de l'infrastructure DCI (composant *DCIManager*) décrit en section 3.3. Ces différents services peuvent être répartis sur différentes machines physiques.

La ***machine de déploiement répartie*** est en charge du processus automatique de déploiement. Elle est constituée de trois composants : le gestionnaire de modèles d'assemblages (composant *AssemblyModelManager*), la fabrique de gestionnaires d'assemblages (composant *AssemblyFactoryManager*) et le gestionnaire d'instances d'assemblages (composant *AssemblyManager)*, décrits en section 3.4.

Les **noeuds d'accueil** (section 3.5) constitués chacun d'un gestionnaire de nœud (composant *NodeManager*), de gestionnaires de serveurs d'applications et de gestionnaires de conteneurs (resp. *ComponentServerManager* et *ContainerManager*) sont présents sur chaque machine où des composants applicatifs doivent être déployés.

*3.2. Les principes de conception des composants DCI*

Comme décrit dans la section 2.3, la spécification CCM ne décrit pas explicitement les relations entre les entités implantant les interfaces de déploiement. Certains chemins de communication entre ces entités étant implicites, l'architecture de l'infrastructure de déploiement CCM n'est pas clairement caractérisée. Pour combler cette lacune, nous avons choisi de concevoir l'infrastructure de déploiement de la plate-forme OpenCCM sous la forme d'un ensemble de composants CORBA pour établir explicitement toutes les dépendances et chemins de communication entre les composants. Ainsi l'architecture de notre infrastructure de déploiement est clairement caractérisée.

D'un point de vue externe, un composant CORBA fournit un ensemble de propriétés configurables ou *attributs* et des ports d'interconnexion avec le monde extérieur. Le CCM propose actuellement quatre types de ports : les interfaces fournies ou *facettes*, les interfaces requises ou *réceptacles*, les *puits* et les *sources d'événements*. Les facettes et les réceptacles permettent d'interconnecter les composants pour des communications par appels de méthodes tandis que les puits et sources permettent des communications asynchrones entre les composants.

L'infrastructure DCI exploite les facettes, réceptacles, sources et puits d'événements : chaque composant propose ainsi des facettes caractérisant les services offerts aux autres composants et des réceptacles exprimant les services requis. Nous distinguons aussi des ports métiers et des ports internes. Les ports métiers capturent les interactions entre composants liées au déploiement d'applications CCM. Pour ces ports, nous réutilisons toutes les interfaces du standard CCM afin de rester compatible avec celui-ci et pouvoir interopérer avec d'autres implantations. Les ports internes quant à eux assurent le maintien de la cohérence de notre architecture DCI. Pour cela, des interfaces propriétaires de gestion des relations entre composants DCI ont été définies, principalement les interfaces suffixées par *Management* et *Registration* décrites dans les sections suivantes. D'un autre côté, les nombreuses associations entre composants utilisent des réceptacles multiples, ce qui offre introspection et contrôle à des outils externes d'administration et de reconfiguration.

### 3.3. Le domaine de déploiement

Le composant *DCIManager* (voir Figure 3) est le gestionnaire central de l'infrastructure DCI. Il se comporte comme un portail d'accès à l'ensemble des fonctions du domaine de déploiement décrites ci-après :

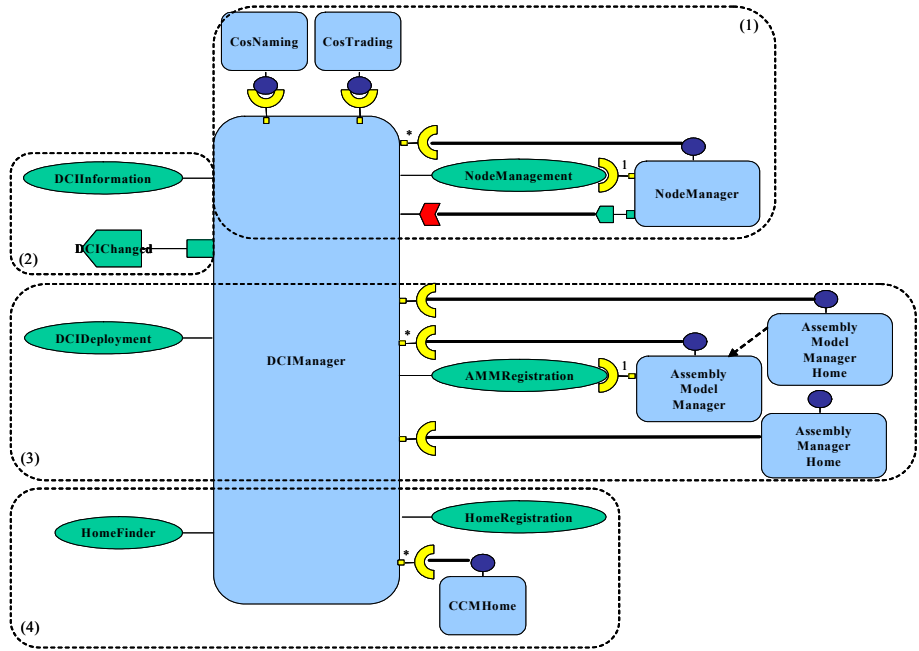

**Figure 3.** *Le composant DCIManager*

(1) La cohérence globale du domaine : le *DCIManager* assure l'utilisation des mêmes instances de *CosNaming* et *CosTrading* par tous les noeuds d'accueil et tous les modèles d'assemblage installés. La facette *NodeManagement* permet aux nœuds de s'enregistrer dynamiquement lors de leurs lancements et de se désenregistrer lors de leurs arrêts.

(2) L'accès aux méta informations sur les caractéristiques matérielles et logicielles (voir section 3.5) des noeuds connectés au domaine de déploiement. La facette *DCIInformation* offre ces informations en XML, ce qui permet à des outils externes de pré-calculer l'assignement des composants sur les noeuds avant leur déploiement.

(3) Le déploiement des assemblages : la facette *DCIDeployment* permet d'installer, de mettre à jour, de rechercher, de lister et de désinstaller des archives (*AssemblyModelManager*), mais aussi de créer, de rechercher, de lister et de détruire des instances d'assemblage (*AssemblyManager*) créées via la maison

*AssemblyManagerHome*. Les modèles d'assemblage sont créés via la maison *AssemblyModelManagerHome* et listés par un réceptacle multiple mis à jour via la facette *AssemblyModelManagerRegistration*.

(4) Le service standard d'annuaire des maisons (*ComponentHomeFinder*) est offert par la facette *HomeFinder*, qui est notamment utilisée par le composant *AssemblyManager* pour retrouver des maisons installées par de précédents assemblages. La liste des maisons enregistrées est stockée dans le réceptacle multiple *CCMHome* et mise à jour par les composants *AssemblyManager* via la facette *HomeRegistration*.

### 3.4. La machine de déploiement répartie

Au cœur de l'infrastructure DCI, la machine de déploiement répartie constituée de trois composantes est en charge de l'automatisation du processus de déploiement des assemblages de composants (voir Figure 4).

Le composant **AssemblyModelManager** représente un modèle d'assemblage installé. A sa création, il reçoit en paramètre l'archive associée à l'assemblage. Il utilise ensuite la facette *AssemblyInstanceManagement* pour en gérer l'ensemble des instances (*AssemblyManager*) : leur création via l'*AssemblyFactoryManager*, leur recherche, leur listage et leur destruction. La liste est maintenue à jour via un réceptacle multiple. Plusieurs *AssemblyModelManager* peuvent partager le même *AssemblyFactoryManager*.

Le composant **AssemblyFactoryManager** offre une facette *AssemblyFactoryExt* étendant l'interface standard du CCM avec quelques méthodes internes aux besoins de l'infrastructure DCI. Ce composant maintient à jour la liste de toutes les instances d'assemblages qu'il démarre via un réceptacle multiple et la facette *AssemblyManagerRegistration*. Ceci autorise l'introspection des assemblages existants, ce qui n'est pas possible avec l'interface standard *AssemblyFactory*.

Le composant **AssemblyManager** modélise une instance d'assemblage déployée et offre deux fonctions principales :

(1) Le contrôle du déploiement grâce à la facette *Assembly* qui permet de lancer les scénarios de déploiement (voir section 2.2) et de repliement d'assemblages. Afin d'assurer le déploiement, le composant *AssemblyManager* interagit avec les services communs du domaine : *DCIInformation* pour retrouver les nœuds disponibles, et *HomeRegistration* / *HomeFinder* / *CosNaming* / *CosTrading* pour les annuaires.

(2) L'introspection de la structure des assemblages déployés via la facette *AssemblyInformation* et les réceptacles multiples vers toutes les instances de composants, maisons, conteneurs, serveurs d'application et nœuds. Cette introspection est fondamentale pour bâtir ensuite des outils d'administration telle que l'interface graphique *OpenCCM Explorer* [OPE 02]. En interne, le composant *AssemblyManager* interprète les descripteurs d'assemblage et coordonne les opérations fournies par les interfaces de déploiement (voir section 2.2).

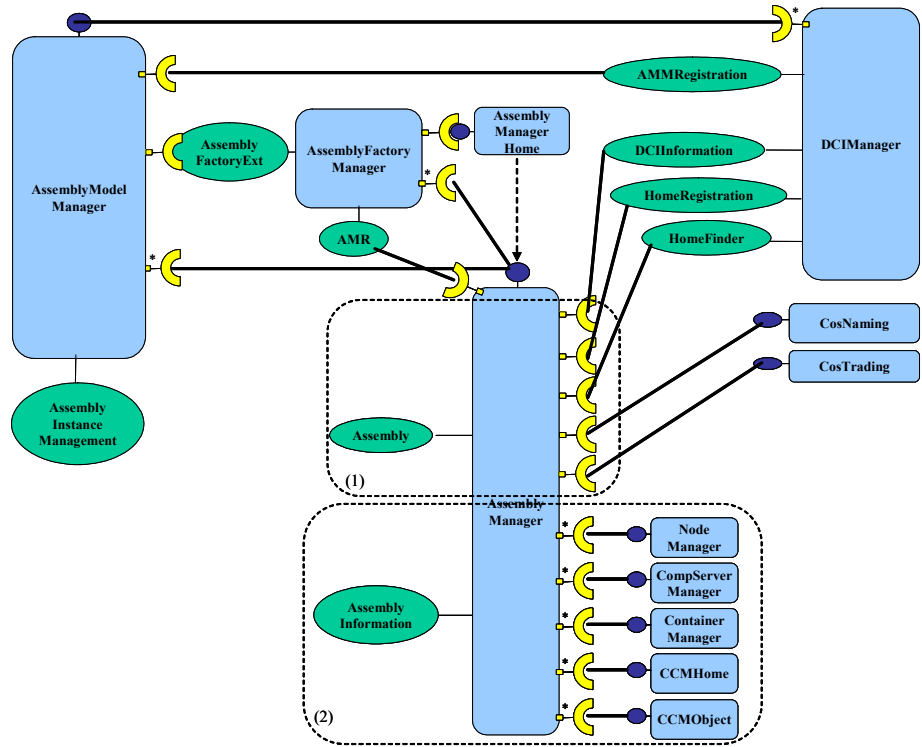

**Figure 4.** *Les composants de la machine de déploiement*

## 3.5. Les nœuds d'accueil

Les composants *NodeManager*, *ComponentServerManager* et *ContainerManager* gèrent un nœud d'accueil. Ils offrent principalement cinq fonctions (voir Figure 5) :

(1) L'accès aux méta informations du nœud est assuré par la facette *NodeInformation* qui fournit en XML les caractéristiques matérielles et logicielles telles que le nombre de processeurs, leur type, l'espace mémoire vive, le système d'exploitation, les logiciels installés, la charge, la bande passante, etc. Leur expression en XML facilite leur structuration, extraction et exploitation pour un assignement automatique ou pour faire de l'équilibrage de charge lors du déploiement. D'autre part, cette facette permet aussi d'obtenir les valeurs fixées pour certaines propriétés du système.

(2) La gestion des archives de composants est assurée par la facette *ExtComponentInstallation* qui offre les opérations pour installer, retrouver, lister et désinstaller les archives de composants sur le nœud d'accueil. Cette interface étend le standard du CCM en y ajoutant l'envoi en mode *push* d'une archive de composant au lieu du seul téléchargement original en mode *pull*.

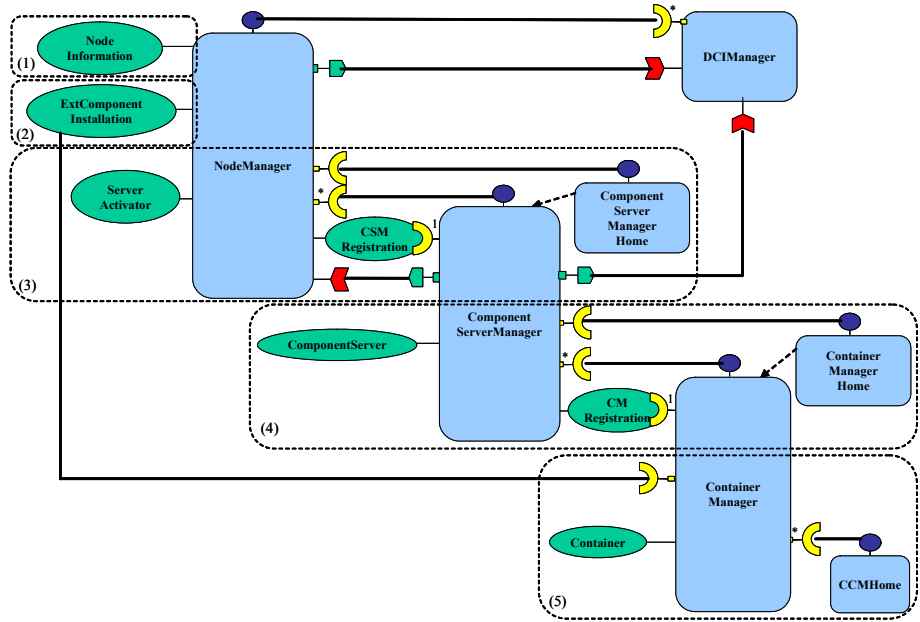

**Figure 5.** *Les composants d'un nœud d'accueil de l'infrastructure OpenCCM DCI*

(3) La gestion des serveurs d'applications (*ComponentServerManager*) est réalisée par quatre ports. La facette *ServerActivator* permet de les créer, lister et détruire. Ils sont créés via la maison *ComponentServerManagerHome* et listés dans un réceptacle multiple mis à jour via la facette *ComponentServerManagerRegistration*.

(4) Chaque instance de *ComponentServerManager* gère un serveur d'applications dans un processus indépendant. La fonction principale d'un serveur d'applications est de gérer un ensemble de conteneurs (*ContainerManager*) par la facette *ComponentServer* offrant des opérations pour créer, lister et détruire les conteneurs. Ceux-ci sont créés via la maison *ContainerManagerHome* et s'exécutent dans le même espace mémoire que le serveur. La liste des conteneurs actifs est stockée dans un réceptacle multiple et mise à jour via la facette *ContainerManagerRegistration*.

(5) Le composant *ContainerManager* fournit la facette *Container* afin d'installer, lister et détruire des maisons de composants. Pour ce faire, il se connecte à la facette *ExtComponentInstallation* fournie par le nœud et en obtient les archives de composants et les implantations binaires afin de les charger en mémoire. La liste des maisons installées (*CCMHome*) est maintenue à jour via un réceptacle multiple.

**4. Conclusion et perspectives**

Le déploiement d'applications réparties à base de composants nécessite la mise en place d'une infrastructure elle-même répartie et dédiée à l'automatisation du processus. Le modèle de composants CORBA spécifie une telle infrastructure (voir section 2.2), dont les lacunes autorisent une large diversité d'implantations. Mais quelques soient les choix d'implantation, ceux-ci doivent répondre à un ensemble de questions architecturales (voir section 2.3). Dans cet article, nous proposons une infrastructure pour le déploiement d'applications à base de composants CORBA elle-même conçue sous cette forme. Par conséquent, elle bénéficie de tous les apports de ce modèle (composants avec ports, conditionnement, déploiement automatisé, conteneurs pour les services non fonctionnels). L'architecture explicite et les différents composants de cette infrastructure ont été spécifiés dans le cadre du projet européen IST COACH et implantés dans notre plate-forme OpenCCM [OPE 03]. L'ensemble fonctionne sur différentes implantations de CORBA (BES, JacORB, ORBacus, Orbix, OpenORB, ZEN), divers systèmes d'exploitation (Linux, Windows, MacOS) et types de machines (station de travail, PDA). Enfin, un complexe système de gestion d'éléments réseaux (Element Management Framework and System – EMF/EMS) réalisé par l'industriel grec Intracom [SET 03], partenaire du projet IST COACH, a permis de valider l'ensemble.

Cette base opérationnelle nous permet de dégager différentes perspectives de travail. Premièrement, notre expérience sur l'application EMF/EMS a montré le besoin d'optimiser le processus de déploiement de grandes applications composées de centaines de binaires à télécharger et de milliers de composants à instancier. Pour cela, nous travaillons sur différentes optimisations au sein du contrôleur de déploiement (le composant *AssemblyManager*). Principalement, nous devons paralléliser l'ordonnancement des tâches élémentaires de processus de déploiement et distribuer le composant *AssemblyManager* sur les différents nœuds accueillant les composants afin de répartir la charge de travail. De plus, un effort important doit être accompli pour minimiser les transferts d'archives et de binaires entre les sites car l'expérience nous a montré que l'installation des composants est la phase la plus consommatrice en temps d'exécution durant le déploiement. Deuxièmement, nous devons aussi distribuer le composant centralisé *DCIManager* sur l'ensemble des sites afin de réaliser le passage à l'échelle et pouvoir adresser le déploiement d'applications sur des grilles de calcul. Troisièmement, l'injection de propriétés non fonctionnelles dans les conteneurs des composants DCI devrait nous permettre de réaliser des déploiements transactionnels, sécurisés et/ou persistants. Enfin, la structuration de l'architecture d'une infrastructure de déploiement n'est pas uniquement nécessaire au déploiement de composants CORBA. Certains éléments de conception de l'architecture OpenCCM DCI devraient pouvoir être réutilisables pour bâtir les infrastructures de déploiement d'autres modèles de composants logiciels répartis comme Fractal, les EJB, les Web Services et plus généralement pour le modèle générique de déploiement et de configuration d'applications réparties à base de composants, récemment standardisé par l'OMG [OMG 03].

## 6. Bibliographie